\documentclass[pra,twocolumn,english,showpacs,floatfix]{revtex4-1}
\usepackage{graphicx}
\usepackage{amssymb}
\usepackage{color}
\usepackage{amsmath}

\graphicspath{{figures_submit/}}

\begin{document}
\title{Thermometry of ultracold fermions by (super)lattice modulation spectroscopy}
\author{Karla Loida, Ameneh Sheikhan, Corinna Kollath}

\affiliation{HISKP, University of Bonn, Nussallee 14-16, 53115 Bonn, Germany}

\begin{abstract}
We theoretically consider non-interacting fermions confined to optical lattices and apply a lattice amplitude modulation that we choose to be either homogeneous or of superlattice geometry. We study the atom excitation rate to higher Bloch bands which can be measured by adiabatic band mapping. We find that the atom excitation rate shows a clear signature of the temperature dependent Fermi distribution in the lowest band of the equilibrium lattice as excitations are quasimomentum-resolved. Based on typical experimental parameters and incorporating a trapping potential, we find that thermometry of one- and two-dimensional systems is within the reach of nowadays experiments. Our scheme is valid down to temperatures of a few percent of the hopping amplitude comparable to the N\'eel temperature in interacting systems.
\end{abstract}

\date{\today}
\maketitle

\section{Introduction}
Ultracold atoms in optical lattices provide a clean realization of various kinds of Hubbard-like models \cite{JakschZoller1998,BlochZwerger2008,Esslinger2010}. These models are well known from the description of solids where they are typically only rough approximations due to the complex structure of materials. Fermionic atoms in optical lattices are of particular interest due to their close analogy to electrons in solids. Experimentalists have succeeded in observing the fermionic Mott-insulator \cite{JoerdensEsslinger2008,SchneiderRosch2008} not long after the first Fermi gas in a three-dimensional optical lattice was characterized \cite{KoehlEsslinger2005}. However, many challenges remain, a major one being further cooling in order to probe interesting regimes such as antiferromagnetic order or unconventional superfluidity \cite{HofstetterLukin2002}. 
Many powerful methods are available to probe ultracold fermionic gases in order to gain precise information on their characteristics. Spectroscopic methods such as radio-frequency, Raman, Bragg or lattice modulation spectroscopy have enabled to probe different characteristic excitations. For instance one can measure single-particle or density spectral functions or collective modes (see \cite{ToermaeSengstock} for a review and references therein). On the other hand, recently great advances have been made towards obtaining real-space information by single-site imaging of fermionic quantum gases \cite{HallerKuhr2015,CheukZwierlein2015,ParsonsGreiner2015}.
However, the temperature determination at low temperatures remains an unresolved issue. Typically, the temperature is measured before and after ramping into the optical lattices. For the harmonically trapped gas, the temperature can be determined from the integrated density profile obtained in a time-of-flight measurement \cite{KetterleZwierlein2008,BlochZwerger2008}. However, this becomes inaccurate at low temperatures, i.e. a few percent of the recoil energy. To determine the temperature within the lattice, entropy conservation during the slow loading process into the lattice is assumed such that the temperature in the lattice can be determined from the initial entropy. This is severely limited by non-adiabatic heating processes caused by the ramping of the lattice or light scattering and fails for in-lattice cooling. Various other schemes to directly determine the temperature inside the optical lattice have been proposed and some have been experimentally tested. For example, the temperature  can be determined from the double occupancy which is sensitive to thermal fluctuations \cite{Koehl2006,JoerdensTroyer2010} for temperatures on the order or above the on-site interaction strength. A temperature measurement has been suggested based on the fluctuation-dissipation theorem and a spatially-resolved density measurement \cite{ZhouHo2011} which requires in-situ resolution in the measurement. Other theoretical proposals include using Raman spectroscopy, transferring the atoms to a third hyperfine state, such that the Raman signal depends on temperature through the Fermi factor \cite{BernierCornaglia2010}, thermometry by light diffraction from atoms in the lattice which results in density-density fluctuations in the scattered intensity \cite{RuostekoskiDeb2009}, or by studying the response to an artificial gauge field \cite{Roscilde2014}. More recently, the temperature of fermions in an anisotropic three-dimensional lattice has been determined from a measurement of the nearest-neighbor spin correlator \cite{ImriskaTroyer2014} and by spin-sensitive Bragg scattering of light \cite{HartHulet2015}. All these methods have their limitations and most cannot be extended into the low-temperature regime of interest. Review \cite{McKayDeMarco2011} gives an overview on some of the thermometry schemes. 

In this work, we suggest a scheme how to directly measure the temperature by means of a time-periodic modulation of the lattice amplitude. Lattice modulation spectroscopy is a well established probe of strongly interacting ultracold atoms in optical lattices as first introduced for bosonic atoms \cite{StoeferleEsslinger2004}. In bosonic systems, energy absorption imprints a characteristic signal in time-of-flight absorption images after sudden switch-off. The energy absorbed by the system is typically estimated from the broadening of the central peak of the momentum distribution. Thus, precise information on the excitation spectrum can be obtained such as a gapped spectrum in the strongly interacting regime \cite{StoeferleEsslinger2004,BlochZwerger2008}. The same procedure is not possible in the case of fermionic systems because here the momentum distribution only depends on temperature by a smearing of the Fermi surface such that very high momentum resolution is necessary in order to extract the energy. However, it was proposed \cite{KollathGiamarchi2006} that a measure of the double occupancy after the lattice modulation gives access to the pairing gap or the interaction energy in the Mott insulator for attractive or repulsive Fermi gases, respectively, as well as information on spin ordering through nearest-neighbor correlations. From the theory side, the double occupancy as a response to lattice amplitude modulation has been extensively studied \cite{HuberRuegg2009,SensarmaDemler2009,MasselToermae2009,KorolyukToermae2010,XuJarrell2011,TokunoGiamarchi2012_0,TokunoGiamarchi2012,DirksFreericks2014}. Experimentally, the former was used to identify the fermionic Mott insulator \cite{JoerdensEsslinger2008} and nearest-neighbor correlations have been probed \cite{GreifEsslinger2011}. The decay of doublons created by the lattice modulation has been measured in order to determine the doublon life-time \cite{StrohmaierDemler2010}. Lattice amplitude modulation has also been used as a spectroscopic probe in order to map out higher Bloch bands in a quasimomentum -resolved fashion \cite{HeinzeSengstock2011} or study interband dynamics \cite{HeinzeBecker2013}.

Motivated by the lack of suitable thermometry schemes we propose in this work to use the lattice modulation spectroscopy in order to directly determine the temperature of non-interacting fermions confined to optical (super)lattices. We consider lattice amplitude modulation of homogeneous and superlattice geometry that create quasimomentum -resolved excitations that enable us to observe signatures of the temperature dependent Fermi distribution. While the commensurate modulation conserves quasimomentum, the superlattice modulation does not which is of interest for many other applications. It injects quasimomentum into the system whenever an excitation is created due to the dimerization of the initial lattice geometry. We show that a temperature determination is possible in both cases for one- and two-dimensional systems at different fillings and within the reach of current experiments. Such a temperature measurement of non-interacting fermions is an important step in order to get also more complex systems under control. In particular, we expect that by an adiabatic connection to many interesting phases our proposal will be widely applicable. Examples of such phases are a weakly interacting Fermi liquid or the weakly interacting Fermi gas in a dimerized superlattice.

This paper is organized as follows. In Section \ref{sec:label1} we introduce the theoretical and experimental framework. We define the equilibrium system and the lattice amplitude modulation of different geometries. We introduce the detection scheme by adiabatic band mapping and the atom excitation rate as our observable in linear response theory. In Section \ref{sec:label2} we develop a multiple band model of the equilibrium system for typical experimental parameters and compute the atom excitation rate to higher bands by a homogeneous lattice modulation. We investigate how this serves as a thermometer in presence of an external trapping potential. In Section \ref{sec:label6} we explore the possibilities of thermometry by a lattice modulation of superlattice geometry applied to both, a homogeneous and a dimerized, equilibrium lattice. In all cases, we consider the experimental feasibility in detail, such as for example requirements on frequency resolution and signal strength. 

\section{Measurement scheme}\label{sec:label1}
\subsection{The unperturbed setup} 
We study a non-interacting gas of fermionic atoms of mass $m$ in two hyperfine states $\sigma=\uparrow,\downarrow$ and with particle number $N=N_{\uparrow}+N_{\downarrow}$ confined to optical lattices.
The Hamiltonian is given by
\begin{eqnarray}
\nonumber H_0&=&\int \text{d}\vec{x} \ \Psi_{\sigma}^{\dag}\left (\vec{x}\right )\bigg [\frac{-\hbar ^2}{2m}\nabla^2 +V_0(\vec{x}) +V_T(\vec{x})-\mu \bigg ]\Psi_{\sigma}\left (\vec{x}\right ),
\end{eqnarray}
where $\Psi_{\sigma}^{(\dag)}\left (\vec{x}\right )$ denotes a fermionic annihilation (creation) operator and $\mu$ is the chemical potential. We approximate the optical lattice potentials by $V_0(\vec{x})=\sum_{i=1}^{3}V_{0,i}\sin^{2}(k_Lx_i)$ with the laser wave number $k_L$ and lattice spacing $a=\pi/k_L$, $x_i$ denotes the directions $x$, $y$ and $z$. Tuning the laser intensity in each direction allows one to choose different lattice strengths $V_{0,i}$ and, consequently, implement different lattice geometries. Three-dimensional (3D) cubic lattices, two-dimensional (2D) pancakes or one-dimensional (1D) tubes of atoms can be realized.
Additionally, we consider an external trapping potential. At the position of the atoms, it can be approximated by a harmonic potential, $V_T(\vec{x})=(m/2)\sum_{i=1}^{3}\omega_{t,i}^{2}x_i^{2}$. This effective trapping potential includes the envelope from the lattice beams.
\subsection{The perturbation}
\begin{figure}
\includegraphics[width=.99\columnwidth,clip=true]{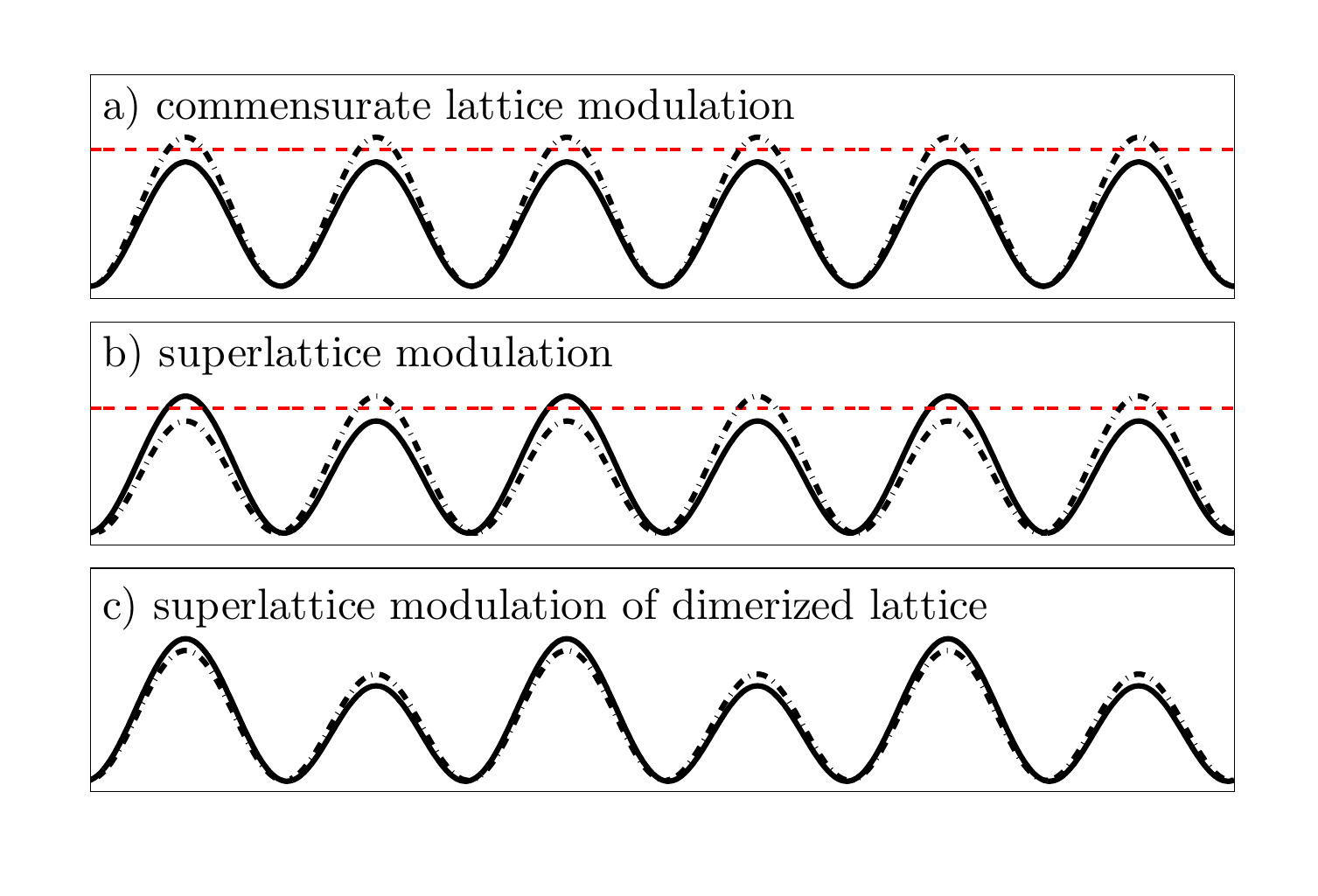}
 \caption{The time-dependent amplitude modulation of the optical lattice between the two configurations indicated by solid and dashed-dotted lines with modulation frequency  $\omega$. We consider different amplitude modulation schemes. In a) the modulation is commensurate with the underlying homogeneous equilibrium lattice and conserves quasimomentum. In b) the modulation is of dimerized (superlattice) geometry and applied to a homogeneous lattice such that quasimomentum is not conserved but quasimomentum $\pi/a=k_L$ is injected into the system. In c) the superlattice modulation is applied to an equilibrium lattice of dimerized geometry such that quasimomentum is conserved. The mean amplitude of the equilibrium lattice is indicated by the horizontal dashed line.}
\label{fig:fig1}
\end{figure}
In order to create controlled excitations which can be used to gain information on the fermionic state, we consider two different perturbations to the optical lattice potential. 
First, we consider standard lattice amplitude modulation \cite{StoeferleEsslinger2004} which is commensurate with the underlying equilibrium lattice as shown in Fig. \ref{fig:fig1} a). The perturbing potential is given by $\delta V(x)=\sin^2(k_Lx)$. Secondly, we consider a superlattice modulation scheme as shown in Fig. \ref{fig:fig1} b). The perturbing potential is then given by $\delta V(x)\approx \sin(k_Lx)$.
Both perturbations are periodically modulated with modulation frequency $\omega$ and small modulation amplitude $A$. In both cases, we choose a modulation of the lattice potential along the direction $x$, regardless of the dimensionality of the underlying equilibrium system. The full lattice potential becomes time-dependent $V(\vec{x},t)= V_0(\vec{x})+A\sin(\omega t) \delta V(x)$. Finally, we also apply the superlattice modulation to a dimerized equilibrium system as depicted in Fig. \ref{fig:fig1} c).
\subsection{The detection scheme}
\begin{figure}
\includegraphics[width=.99\columnwidth,clip=true]{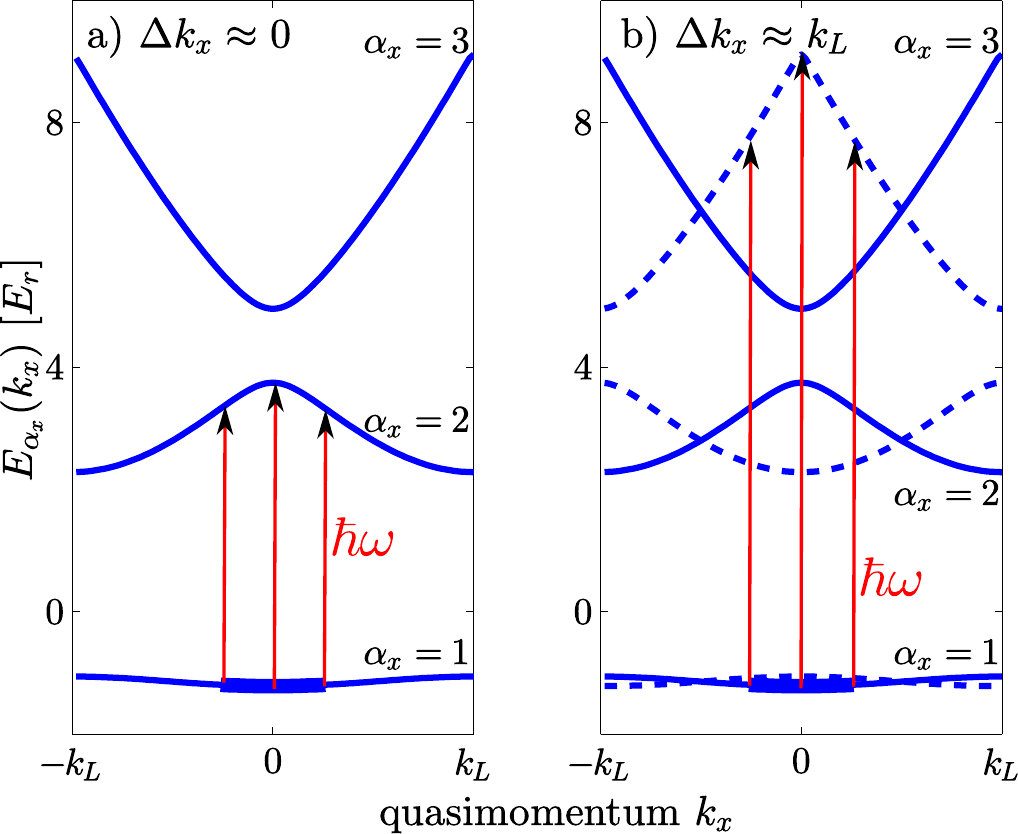}
 \caption{The three lowest Bloch bands of a homogeneous optical lattice along $x$-direction and possible excitations by lattice amplitude modulation. We consider $V_{0,x}=7E_r$  in units of the atomic recoil energy $E_r=(\hbar k_L)^2/2m$. In the case of a commensurate modulation a) the perturbing frequency is chosen such that atoms are transferred from the lowest band $\alpha_x=1$ to the first excited band $\alpha_x'=2$  with conserved quasimomentum $k_x$. In the case of a superlattice modulation b) the perturbing frequency is chosen such that atoms are transferred from the lowest band $\alpha_x=1$ at quasimomentum $k_x$ to the second excited excited band $\alpha_x'=3$ at quasimomentum $k_x+k_L$ which is equivalent to exciting to the shifted band $E_{\alpha_x=3}(k_x+k_L)$ (dashed line) at same quasimomentum $k_x$. The shifted bands $E_{\alpha_x=1}(k_x+k_L)$ and $E_{\alpha_x=2}(k_x+k_L)$ are also depicted by dashed lines. In both cases the minimum and maximum value of possible excitation energies are related to the initial filling of the lowest band indicated by a thicker line that corresponds to quarter-filling in this figure. The minimum possible excitation energy is indicated by red arrows.}
\label{fig:fig2}
\end{figure}
Determining the energy absorbed by fermionic systems from the momentum distribution in time-of-flight images is much harder than for bosons. The momentum distribution is step like due to fermionic statistics and not as sensitive to heating. In the following, we propose a simple measurement scheme of the response of the atoms to the lattice amplitude modulation. The concept relies on the excitation to higher Bloch bands combined with the adiabatic band mapping technique.\\
We assume the lattice potential to be sufficiently deep such that the Bloch bands are well separated and initially only the lowest band is occupied. We choose lattice modulation frequencies such that we excite from the lowest band $\alpha_x=1$ to the first excited band $\alpha_x'=2$ or to the second excited band $\alpha_x'=3$ along $x$-direction. Possible excitations of the modulation along $x$-direction are sketched in Fig. \ref{fig:fig2}. In the case of commensurate lattice modulation the transition matrix elements are non-zero for conserved quasimomentum $\Delta k\approx 0$ such that excitations occur vertically in quasimomentum space.
 We concentrate on excitations to the first excited band $\alpha_x'=2$. We comment on this choice in Sec. \ref{sec:label3} in more detail. For each excitation energy $\hbar \omega$ only one initial quasimomentum pair $\pm k_x$ is on resonance due to the inverted and stretched form of the $\alpha_x'=2$ band. Thus, a quasimomentum resolved excitation is ensured and can be detected by a measurement of the transferred band occupation $ \langle n_{\omega}^{\alpha_x'}(t) \rangle$ of the excited band $\alpha_x'$ without the necessity of quasimomentum resolution within the band. Possible excitations strongly depend on the filling of the lowest band such that the Fermi occupation which depends on temperature will be reflected in the response to the lattice modulation. Moreover, the Fermi dependence of the response will be stretched in energy space compared to the Fermi distribution in the lowest band such that the required frequency resolution needed in experiment to resolve the Fermi tail is reduced. Band inversion alone leads to a stretching of the Fermi distribution by a factor two which is strongly enhanced because the excited band is much broader in energy than the lowest band. 

In the case of superlattice modulation spectroscopy applied to a homogeneous lattice the transition matrix elements are non-zero for a quasimomentum transfer of  $\Delta k_x\approx k_L$ such that excitations occur vertically in quasimomentum space only if we shift the excited band in quasimomentum space by $\Delta k_x=k_L$. We now concentrate on excitations to band $\alpha_x'=3$ in order to obtain a similar band structure geometry as for the commensurate modulation scheme as shown in Fig. \ref{fig:fig2} b). Note, that the required experimental frequency resolution is further lowered compared to the commensurate modulation because of the larger width of the second excited band compared to the first excited band.

In the case of the superlattice modulation spectroscopy applied to a dimerized lattice excitations occur with zero quasimomentum transfer. Dimerization leads to a folding of the first Brillouin zone such that the lowest band is split into two bands. Excitations occur from the lower to the upper band with zero quasimomentum transfer in the reduced zone scheme which corresponds to a quasimomentum transfer  $\Delta k_x\approx k_L$ in the extended zone scheme.

 Note that Ref.~\cite{BernierCornaglia2010} also suggests the implementation of a finite momentum transfer in order to increase the frequency resolution of the measured temperature dependence. However, the experimental realization proposed uses a Raman process which transfers a portion of the atoms in the lattice to a different hyperfine state. The proposal uses two additional Raman beams. In contrast, in our proposal the lattice potential itself is used in order to generate the excitations, also at finite quasimomentum, such that it might be easier to integrate in certain experimental setups.

\subsection{Linear response theory}
We use linear response theory in order to determine the rate of excitations $\partial_t\langle n_{\omega}^{\alpha'}\rangle$ for the atoms into higher bands $\alpha'$. Linear response is a common method used in order to describe the response of a many-body system to a perturbation  \cite{CohenTannoudjiLaloe}. It has been proved a very powerful tool for the description of lattice shaking experiments \cite{KollathSchollwoeck2006}. Typically, the time-evolution of the energy during the application of a weak perturbation which is close to resonance of some excitations shows initially a quadratic rise, which after a short time goes over to a linear increase. At long times a saturation of the absorbed energy sets in. The slope of the intermediate linear rise can be related to the atom excitation rate determined in linear response theory. 
For a small perturbing amplitude $A$ the rate of excitations within linear response theory is given by
\begin{eqnarray}
\nonumber \frac{1}{\vert A\vert^2}\partial_t\langle n_{\omega}^{\alpha'}\rangle&=&\frac{1}{2\hbar}\frac{\pi}{Z}\left (1-e^{-\beta \hbar \omega}\right)\sum_{n,m}\Big \{\vert \langle m\vert O \vert n\rangle\vert^2\\
&\times&e^{-\beta E_n}\delta \left(\hbar \omega+E_n-E_m\right)\Big \},
\label{eqn:1}
\end{eqnarray}
where $Z=\sum_n\exp(-\beta E_n)$  is the partition sum, $\beta =1/(k_BT)$ is the inverse temperature, $\vert n \rangle $ is an eigenstate of the unperturbed system $H_0$ with corresponding eigenenergy $E_n$ and  $O=\int \text{d}\vec{x} \ \Psi_{\sigma}^{\dag}\left (\vec{x}\right ) \delta V(x)\Psi_{\sigma}\left (\vec{x}\right )$ is the perturbing part of the Hamiltonian. This formula is known as the spectral representation of the response and allows for an intuitive interpretation. An excitation is created whenever the perturbation $O$ couples many-body state $\vert n\rangle$ to many-body state $\vert m\rangle$ and conservation of energy, $\hbar\omega=E_m-E_n$, is satisfied. 

Note, that the application of linear response to the considered situation of non-interacting fermions might seem counter-intuitive at a first sight because exact momentum-resolution of the sinusoidal perturbation would lead to a resonant coupling of two discrete levels and induce clean Rabi-oscillations. However, due to the trapping potential, weak interactions and a finite perturbation time, a group of states is excited by the perturbation yielding a linear rise in energy for sufficiently large times \cite{CohenTannoudjiLaloe}. We have verified the applicability of linear response theory at an example case using a time-dependent density-matrix renormalization group study. 

\section{Thermometry by commensurate lattice modulation spectroscopy}\label{sec:label2}

In this section we investigate the response of the system to a commensurate lattice modulation at frequencies corresponding to higher-band excitations. We focus on excitations to the first excited band $\alpha_x'=2$ and study the response of the homogeneous system as well as the trapped system within the local density approximation (LDA).  We show how the Fermi distribution of the atoms in the lowest band imprints a clear signal in the response such that thermometry is possible for one- and two-dimensional equilibrium systems.

\subsection{The two-band tight-binding model}\label{sec:label3}
\begin{figure}
\includegraphics[width=.99\columnwidth,clip=true]{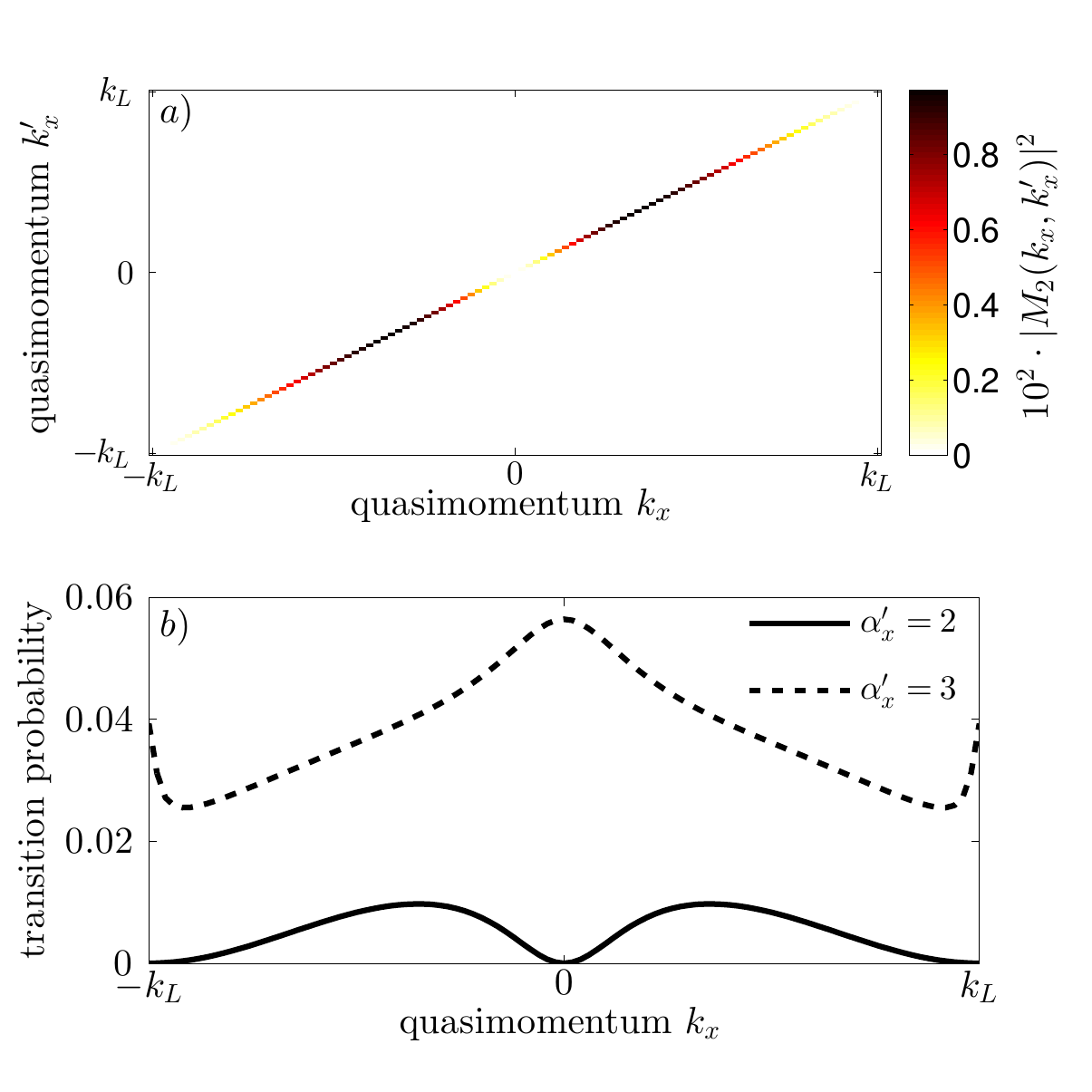}
 \caption{a) The transition probability $\vert M_{\alpha_x'=2}(k_x,k_x')\vert^2$ for exciting atoms at quasimomentum $k_x$ from the lowest band $\alpha=1$ to  the first  excited band $\alpha'=2$  for all $k_x'$ by commensurate lattice modulation. b) The transition probability $\vert M_{\alpha_x'}(k_x)\vert^2$ for exciting atoms to the first $\alpha'=2$ or second $\alpha'=3$ excited band at quasimomentum $k_x'=k_x$ by commensurate lattice modulation. We consider a lattice depth of $V_{0,x}=7E_r$. }
\label{fig:fig3}
\end{figure}
In this subsection we derive a tight-binding description of the considered situation. We assume a sufficiently deep optical lattice potential such that the emerging lowest Bloch bands are well separated. Only the lowest band $\alpha=1$ and one excited band $\alpha_x'$ are important in our considerations and thus taken into account.
We consider an atom trapped in the $d$-dimensional periodic potential $V_0(\vec{x})$ in the absence of an additional external trapping potential. We compute the eigenfunctions which are the Bloch functions along the different directions $\phi_{\alpha_{x_i}} ^{k_{x_i}}(x_i)$ with corresponding eigenvalues (the Bloch bands) $E_{\alpha_{x_i}}(k_{x_i})$ where $\alpha_{x_i}$ denotes the band index and the quasimomentum $k_{x_i}$ lies within the first Brillouin zone $]-k_L,k_L]$. The full spectrum is given by $E_{\alpha}(\vec{k})=\sum_{i=1}^{d}E_{\alpha_{x_i}}(k_{x_i})$ as we consider non-interacting fermions for which the potential is separable. The index $\alpha=\{\alpha_x,\alpha_y,\alpha_z\}$ labels the band. Note, that the Bloch functions factorize $\phi_{\alpha}^{\vec{k}}(\vec{x})=\prod_{i=1}^{d}\phi_{\alpha_{x_i}}^{k_{x_i}}(x_i)$ in the tight-binding approximation neglecting the coupling to other Bloch bands and the overlap of different sites. The unperturbed Hamiltonian (for $V_T=0$) becomes
\begin{eqnarray}
H_0&\approx&\sum_{\alpha,\vec{k},\sigma}\left [E_{\alpha}(\vec{k})-\mu \right ]c_{\alpha\vec{k}\sigma}^{\dag}c_{\alpha\vec{k}\sigma},
\label{eqn:2}
\end{eqnarray}
where $c_{\alpha\vec{k}\sigma}^{(\dag)}$ are the fermionic annihilation (creation) operators. \\
In order to construct the perturbing Hamiltonian, we need to determine the transition matrix elements of the perturbing potential $\delta V(x)$ in the chosen Bloch basis. Initially, only the lowest band $\alpha=1\equiv \{\alpha_x=1,\alpha_y=1,\alpha_z=1\}$ is occupied. Thus, we are only interested in the transition matrix elements given by
\begin{eqnarray}
\nonumber M_{(\alpha=1,\vec{k})\rightarrow (\alpha',\vec{k}')}&=& \frac{1}{\Omega_x}\int_{x_{\text{min}}}^{x_{\text{max}}}\text{d}x\ \phi_{\alpha_x'}^{k_x'*}(x)\delta V(x) \phi_{\alpha_x=1}^{k_x}(x)\\
&\times& \prod_{i\neq 1}\delta_{\alpha_{x_i}',\alpha_{x_i}=1}\ \delta_{k_{x_i}',k_{x_i}},\label{eqn:3}
\end{eqnarray}
where $\Omega_x=(L-1)a$ is the system size in $x$-direction with $L$ the number of lattice sites and $x_{\text{min}}=-(L/2-1)a$ and $x_{\text{max}}=La/2$. We used that the Bloch functions are orthogonal, i.e. $(1/\Omega_{x_i})\int_{x_{i,\text{min}}}^{x_{i,\text{max}}}\text{d}x_i\ \phi_{\alpha_{x_i}'}^{k_{x_i}'*}(x_i) \phi_{\alpha_{x_i}}^{k_{x_i}}(x_i)=\delta_{\alpha_{x_i}',\alpha_{x_i}}\delta_{k_{x_i}',k_{x_i}}$. We insert the perturbing potential $\delta V(x)=\sin^{2}(k_Lx)$ and find that the perturbation, to a good approximation, only couples momenta $\vec{k}$ and $\vec{k}'\approx \vec{k}$ as shown in Fig. \ref{fig:fig3} a). All other matrix elements for $\Delta \vec{k}\neq 0$ are strongly suppressed. This justifies the approximation that the quasimomentum is conserved by the commensurate lattice modulation. We approximate
\begin{eqnarray}
\nonumber M_{(\alpha=1,\vec{k})\rightarrow (\alpha',\vec{k}')}&=&M_{ \alpha_x'}(k_x,k_x') \prod_{i\neq 1}\delta_{k_{x_i}',k_{x_i}}\\
 \nonumber &\approx&M_{ \alpha_x'}(k_x)\delta_{\vec{k},\vec{k}'},
\end{eqnarray}
where excitations occur to the band $\alpha'=\{\alpha_x',\alpha_y=1,\alpha_z=1\}$.
\\In the tight-binding approximation the perturbing part of the Hamiltonian becomes
\begin{eqnarray}
\nonumber H_{\text{pert}}&\approx&A\sin(\omega t)O,\\
O&=&\sum_{\vec{k},\sigma}M_{\alpha_x'}(k_x)\left(c_{\alpha'\vec{k}\sigma}^{\dag}c_{\alpha=1\vec{k}\sigma}+\text{h.c.}\right)
\label{eqn:4}
\end{eqnarray}
The probability $\vert M_{ \alpha_x'}(k_x) \vert^{2}$ of exciting atoms from the lowest band to the band $\alpha_x'=2$ or $\alpha_x'=3$ with $\Delta k_x=0$ is shown in Fig. \ref{fig:fig3} b). Excitations to the first excited band are prohibited at the center $k_x=0$ and the border $k_x=k_L$ of the Brillouin zone due to the symmetry of the Bloch functions. The maximum probability lies in between. This is advantageous as this region coincides with the region of interest around the Fermi surface of a half-filled lowest band such that the temperature dependence of the Fermi surface can be probed by the response. Moreover, the amplitude is sufficiently strong such that enough atoms for a detectable signal are excited at small perturbing amplitudes and reasonable perturbing times. We supply numbers for an example in Section \ref{sec:label4} in presence of an external trap. In contrast, excitations along $x$ to the second excited band $\alpha'=3$ are non-zero for all momenta and the amplitude is about three times larger.
 The transition probability $\vert M_{ 3}(k_x) \vert^{2}$ is maximum at $k_x=0$ and $k_x=k_L$. The corresponding contributions dominate the response at the corresponding frequencies if considering the full range of excitations.
This makes it more favorable to use excitations to  $\alpha'=2$ for a temperature measurement at intermediate lattice heights.
 Note that stronger transition amplitudes to $\alpha'=3$ than to  $\alpha'=2$ are in agreement with the expectations for a large lattice height where lattice wells almost decouple. Approximating each well by a harmonic oscillator, the modulation corresponds to a frequency modulation of the quantum harmonic oscillator, $\tilde{\omega}_{HO}=\omega_{HO} (1+\rho)$, where $\rho$ is a small parameter. This modulation couples to the second excited band, but transitions to the first excited band are prohibited by symmetry. We confirm that  $\vert M_{ 2}(k_x) \vert^{2}$ decreases for increasing lattice depths which corresponds to approaching the harmonic oscillator limit in a single well. 

\subsection{The response of the homogeneous system in one dimension}
\begin{figure}
\includegraphics[width=.99\columnwidth,clip=true]{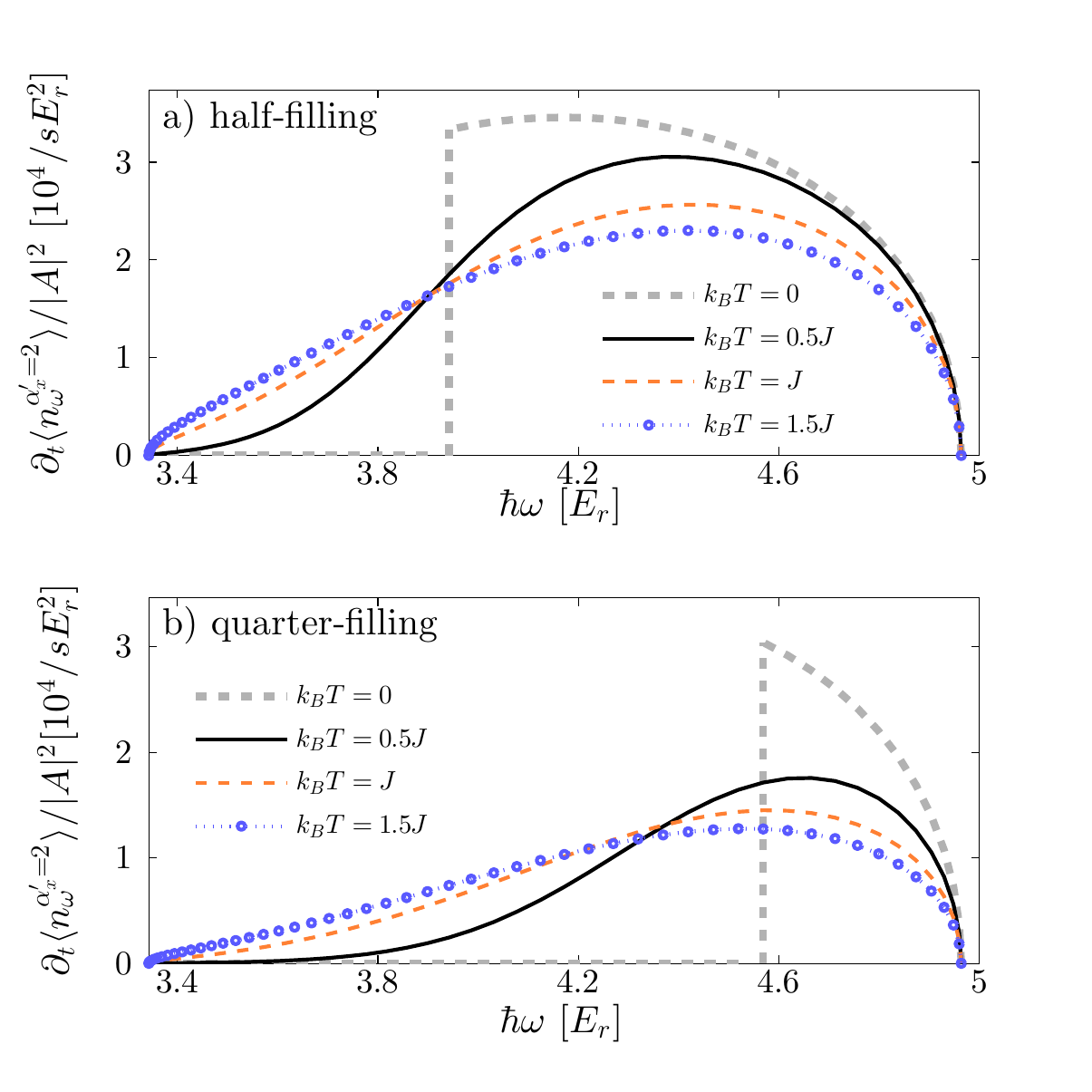}
 \caption{Excitation rate of the atom number to the first excited band $\alpha_x=2$ by the commensurate lattice modulation spectroscopy of the homogeneous system for a) half-filling and b) quarter-filling for temperatures $k_BT=0$, $0.5$J, $1$J and $1.5$J.}
\label{fig:fig4}
\end{figure}
We start our discussion of results by the simplest system which consists of homogeneous 1D tubes oriented along $x$-direction. Assuming that only $\alpha_x=1$ is initially occupied, it can be shown that the response function Eq.~\eqref{eqn:1} in the continuum limit becomes
\begin{eqnarray}
\nonumber&&\frac{1}{\vert A\vert^2}\partial_t\langle n_{\omega}^{\alpha_x'}\rangle^{1D}=\frac{\pi L}{\hbar k_L} \int_0^{k_L}\text{d}k_x\ \vert M_{ \alpha_x'}(k_x)\vert^2\\
&&\times f\left [E_{\alpha_x=1}(k_x)-\mu)\right ] \delta \left \{ \hbar \omega-\left[E_{ \alpha_x'}(k_x)-E_{\alpha_x=1}(k_x)\right] \right\} .
\label{eqn:5a}
\end{eqnarray}
This expression strongly depends on temperature through the Fermi function $f(E-\mu)=1/\{1+\exp\left [(E-\mu)/k_BT\right ]\}$. We show the results for excitations to the first excited band $\alpha_x'=2$ for different fillings and temperatures in Fig. \ref{fig:fig4}. Here we consider $^{40}K$-atoms and a laser wave length of $\lambda=1064$nm. All energies are conveniently expressed in units of the atomic recoil energy $E_r=h^2/2m\lambda^2$. We use these parameters throughout the remainder of this work. Temperatures we express in units of the tunneling amplitude of atoms in the lowest band between adjacent sites of the optical lattice $J=(E_{\alpha_x=1,\text{max}}-E_{\alpha_x=1,\text{min}})/4\approx 0.04Er$. The zero-temperature response vanishes for excitation energies below the minimum possible excitation energy. At zero temperature, the minimum excitation energy corresponds to the excitation energy of an atom located at the Fermi surface (cf. Fig. \ref{fig:fig2} a) ). Smaller energies correspond to states located above the Fermi surface which are empty. At higher excitation energies the response is non-zero because atoms below the Fermi surface can be addressed. The maximum possible excitation energy is always given by the energy needed to excite an atom located at $k_x=0$. At finite temperatures atoms get thermally excited around the Fermi surface. The Fermi step softens according to the Fermi function such that the excitations depend on the Fermi distribution in the lowest band. The minimum excitation energy is no longer sharp but the Fermi dependence is reflected in the response as a function of excitation energy.  We compare half-filling in Fig. \ref{fig:fig4}a) and quarter-filling in Fig. \ref{fig:fig4}b) for temperatures $k_BT=0$, $0.5J$, $J$ and $1.5J$. For both fillings the temperature dependence of the response is clearly visible. At higher temperature the Fermi step smears out more which results in a broadening of the low frequency tail, in particular at low fillings. The possible broadening is limited by the upper band edge, limiting the resolution at high temperatures. The position of the Fermi tail shifts to smaller energies for larger fillings due to the dependence of the excitation energy $\hbar \omega$ on the quasimomentum $k_x$. Furthermore, the Fermi tail is broadened in energy due to the enlarged bandwidth of the first excited band with respect to the lowest band. The different response functions can be clearly distinguished for the displayed temperatures.

\subsection{The response of the trapped system in one dimension}\label{sec:label4}
\begin{figure}
\includegraphics[width=.99\columnwidth,clip=true]{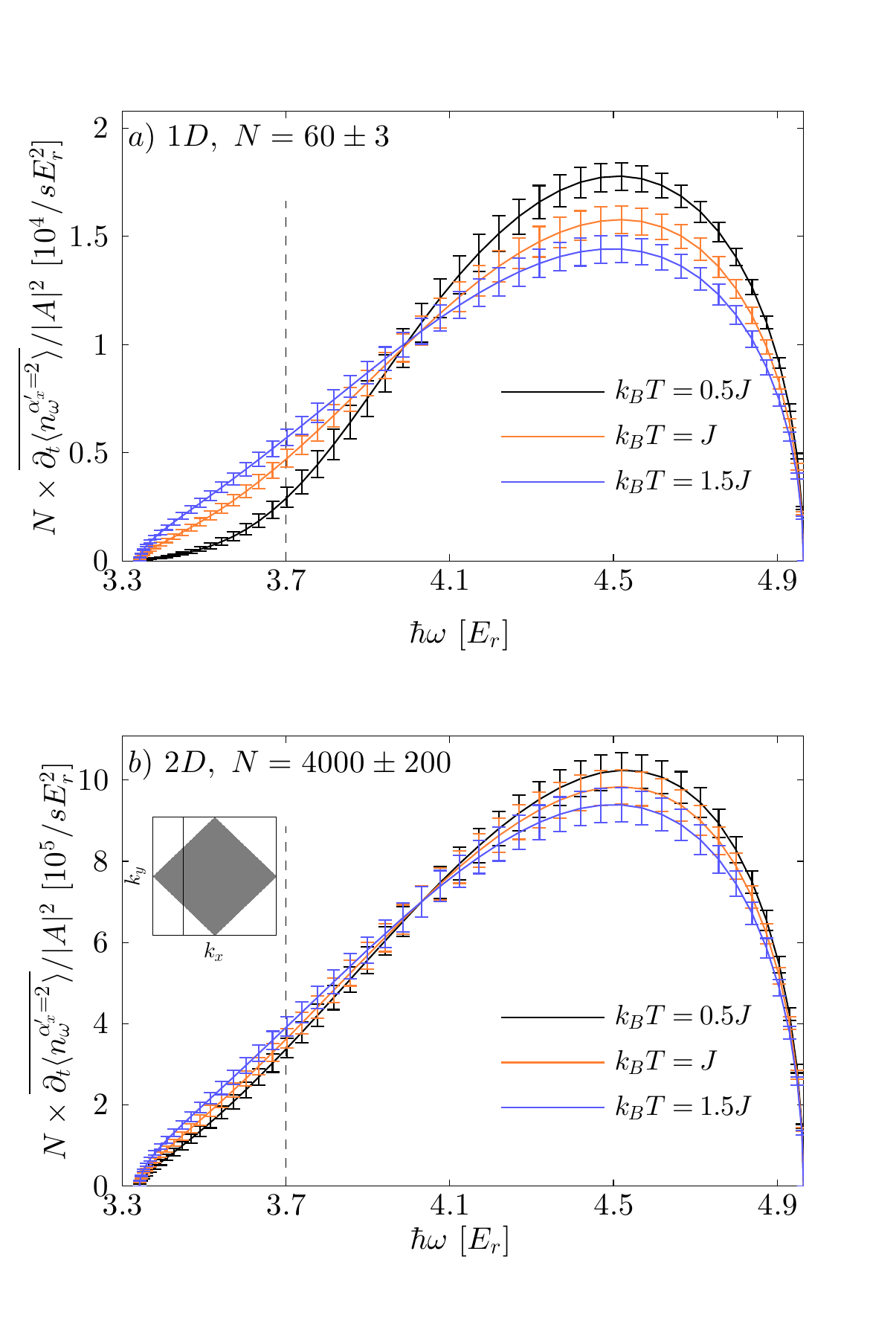}
 \caption{Excitation rate of the atom number to the first excited band $\alpha_x=2$ of the commensurate lattice modulation spectroscopy of the trapped system. In a) the underlying equilibrium lattice is one-dimensional with atom number $N=60$. In b) the underlying equilibrium lattice is two-dimensional with atom number $N=4000$. In both cases, the response is shown for temperatures $k_BT=0.5$J, $1$J and $1.5$J with error bars corresponding to $5\%$ uncertainty on the initial atom number $N$ and $k_BT=0$ is shown in a) . The vertical lines indicate $\hbar \omega=3.7E_r$ for which we determine the mean number of excited atoms (see main text). The inset in b) shows the first Brillouin zone of the square lattice system. At zero temperature and half filling all states below the Fermi surface are occupied (gray) and all states above the Fermi surface are empty (white). At a fixed $k_x$ (vertical black line) different $k_y$ contribute such that the response is a superposition of constant contributions in the bulk and temperature dependence mainly comes into play near the Fermi surface.}
\label{fig:fig5}
\end{figure}
As we have seen that the response depends strongly on the filling, we continue our discussion by adding an external harmonic trapping potential as present in many experimental setups. We treat the trap within the LDA. In appendix \ref{app:comparison} a detailed comparison of exact calculations and the LDA analysis is performed which supports the validity of the used approximation. The space-dependence of the trap $V_T(x)=V_t(x/a)^2$ is incorporated into the chemical potential. The chemical potential thus varies when moving through the trap, i.e. $\mu(x)=\mu_0-V_T(x)$, where $\mu_0$ is the chemical potential at the center of the trap. We obtain the mean response $\overline{\partial_t\langle n_{\omega}^{\alpha_x'}\rangle^{1D}}/\vert A\vert^2$ of the trapped system by summing the homogeneous response per lattice site $\partial_t\langle n_{\omega}^{\alpha_x'}\rangle^{1D}/(\vert A\vert^2L)$ over all chemical potential present in the trap with the correct weight and normalizing by the total number of atoms. The mean response of the trapped system becomes
\begin{eqnarray}
&&\frac{\overline{\partial_t\langle n_{\omega}^{\alpha_x'}\rangle^{1D}}}{\vert A\vert^2}=\frac{1}{NL\vert A\vert^2}\sqrt{\frac{2J}{V_t}}\int_{-\infty}^{\overline{\mu}_0}\text{d}\bar \mu \frac{\partial_t\langle n_{\omega}^{\alpha_x'}\rangle^{1D}}{\sqrt{\bar \mu_0 - \bar \mu}}
\label{eqn:5}
\end{eqnarray}
with $\bar \mu=\mu/2J$ and $V_t=(m/2)\omega_{t,x}^2a^2$. The chemical potential in the trap center $\mu_0$ is determined by the total atom number $N$. The mean response for an initial atom number $N=60$ and a trapping frequency $\omega_{t,x}=2\pi\times 24.5$Hz$=5.6\times 10^{-3}E_r$ in a 1D system is shown in Fig. \ref{fig:fig5} a) for temperatures  $k_BT=0.5$J, $1$J and $1.5$J. Error bars correspond to $5\%$ uncertainty on the initial atom number $N$. The results resemble the response of the homogeneous system and a temperature dependence is clearly visible. However, the Fermi tail is less distinct as we sum over different fillings present in the trap. In order to resolve the Fermi tail and distinguish the curves of different temperatures we estimate from the horizontal distance of the curves in Fig. \ref{fig:fig5} a) that a frequency resolution $\Delta (\hbar \omega) \approx \pm 0.025E_r$ is required in experiment which corresponds to a perturbing time $t=2\pi/\Delta \omega\approx 9$ms. We determine the mean number of atoms $ \langle n_{\omega}^{\alpha_x'}(t) \rangle$ excited during the time $t$ which is the observable in experiments. The mean number of excited atoms in $d$-dimensional systems is given by
\begin{eqnarray}
\langle n_{\omega}^{\alpha_x'}(t) \rangle&=&  N \times \overline{\partial_t\langle n_{\omega}^{\alpha_x'}\rangle^{dD}}\times t .
\label{eqn:6}
\end{eqnarray}
We choose a sufficiently small amplitude modulation of $A=0.05V_{0,x}$ where $V_{0,x}=7E_r$ and consider the point $\hbar \omega=3.7E_r$ as an example, indicated by a vertical line in Fig. \ref{fig:fig5} a). Using Eq. \eqref{eqn:6} we determine the mean number of excited atoms $\langle n_{\omega}^{\alpha_x'=2}(t) \rangle$ to lie between $3$ and $7$ for temperatures $k_BT=0.5$J, $1$J and $1.5$J. Considering $\sim 100$ parallel 1D tubes of roughly equal filling in experiment, this gives a difference in number of excited atoms of $\sim 150$ between the chosen temperatures which is measurable by current experimental means.

\subsection{The response of the trapped system in higher dimensions}\label{sec:label5}

We may also probe higher-dimensional equilibrium systems by lattice amplitude modulation along one direction.
The response of $d$-dimensional lattices becomes
\begin{eqnarray}
\nonumber &&\frac{\partial_t\langle n_{\omega}^{\alpha'}\rangle^{dD}}{\vert A\vert^2}=\frac{\pi}{\hbar}\left (\frac{L}{2k_L}\right)^d \nonumber \int_{BZ}\text{d}^dk\ \vert M_{ \alpha_x'}(k_x)\vert^2\\
\nonumber && \times f\left [E_1(\vec{k})-\mu)\right ]  \delta \left \{\hbar \omega-\left[E_{ \alpha_x'}(k_x)-E_{\alpha_x=1}(k_x)\right] \right \} ,
\end{eqnarray}
where the integral runs over the first Brillouin zone (BZ).
Due to the modulation along the $x$-direction only, the resonance condition in this expression is set by the change of the energy in $x$-direction, $E_{ \alpha_x'}(k_x)-E_{\alpha_x=1}(k_x)$, and the matrix element depends only on $k_x$. The difference to the system of 1D tubes enters in the initial Fermi distribution where the energy of the $d$-dimensional system $E_1(\vec{k})$ occurs.
For $d>1$ the signal remains $k_x$ resolved, however, many different $k_y$-quasimomenta contribute at each fixed $k_x$. These points have different location in $\vec{k}$-space with respect to the Fermi surface such that we detect a superposition of different points in the Fermi distribution. We illustrate this for half-filling in a 2D system in the inset of Fig. \ref{fig:fig5} b). Consider an intermediate $k_x$-value as indicated by the vertical line. Along this line a temperature dependence will only show up close to the Fermi surface whereas the bulk dominantly contributes with a constant value to the response. Consequently, the effect of temperature in the response is less pronounced. However, in 2D a temperature measurement is still possible but requires better frequency resolution in experiment than the 1D case. Within LDA we obtain
\begin{eqnarray}
\frac{\overline{\partial_t\langle n_{\omega }^{\alpha'}\rangle^{2D}}}{\vert A\vert^2}&=&\frac{\pi}{NL^2\vert A\vert^2}\left (\frac{4J}{\overline{V_t}}\right )\int_{-\infty}^{\overline{\mu}_0}\text{d}\bar \mu \ \partial_t\langle n_{\omega}^{\alpha'}\rangle^{2D}(\bar \mu),
\label{eqn:7}
\end{eqnarray}
where  $\bar \mu=\mu/4J$ and $\overline{V_t}=(m/2)\sqrt{\omega_{t,x}^2\omega_{t,y}^2}a^2$.
The result for trapping frequencies $\omega_{t,x}=2\pi \times24.5$Hz$=5.6\times 10^{-3}E_r$ and $\omega_{t,y}=2\pi \times29.7$Hz$=6.7\times 10^{-3}E_r$ and an initial atom number $N=4000$ are shown in Fig. \ref{fig:fig5} b) for temperatures  $k_BT=0.5$J, $1$J and $1.5$J. Error bars correspond to $5\%$ uncertainty on the initial atom number. In order to resolve the Fermi tail and distinguish the curves of different temperature a frequency resolution $\Delta (\hbar \omega) \approx \pm 0.01E_r$ is required in experiment which corresponds to a perturbing time $t\approx 23$ms for the same system parameters as in Section \ref{sec:label4}. We choose a small amplitude modulation of $A=0.02V_0$ and consider the point $\hbar \omega=3.7E_r$ as an example, indicated by a vertical line in Fig. \ref{fig:fig5} b). Using Eq. \eqref{eqn:6} we determine the mean number of excited atoms  $\langle n_{\omega}^{\alpha'=2} (t) \rangle$ to lie between $150$ and $180$ for temperatures $k_BT=0.5$J, $1$J and $1.5$J. This gives a difference in number of excited atoms of $\sim 10$ between the chosen temperatures. This may be demanding to measure but achievable with current experimental techniques. \\
Note that applying the perturbation along several directions does not help in order to gain on resolution. The modulation decouples into the different directions, for example a lattice modulation along two directions $\delta V(\vec{x})=\delta V(x)+\delta V(y)$, such that it has the effect of two 1D perturbations along $x$ and $y$, respectively. As one has the contributions of both directions, the total number of excited atoms is only increased by a factor of two for a completely isotropic setup. We do not benefit from this factor as we are limited in minimum time by frequency resolution.

\section{Thermometry by superlattice modulation spectroscopy}\label{sec:label6}

In this section we study the possibilities of thermometry by superlattice modulation spectroscopy. We consider two different equilibrium situations. First, we consider the homogeneous optical lattice system, where all wells are equal and of spacing $a=\lambda/2$, subjected to a superlattice modulation. In this situation the symmetry of the perturbation is different with respect to the equilibrium system. The periodicity in space is doubled and now $2a$ whereas the Brillouin zone has half the size with respect to the equilibrium lattice. Secondly, we consider a dimerized equilibrium lattice such that the superlattice modulation scheme has the same symmetry as the underlying lattice geometry.\\
In the first case, we discuss higher-band excitations along the lines of the previous Section \ref{sec:label2} where we now choose to consider the excitations to the second excited band $\alpha'=3$.
The lattice geometry becomes dimerized which strongly affects the nature of excitations, namely excitations with quasimomentum  $k_L$ are created.\\In the second case, we probe a dimerized equilibrium system (superlattice geometry) such that the superlattice modulation is commensurate with the initial geometry [cf. Fig. \ref{fig:fig1} c)] and conserves quasimomentum. In this case, dimerization leads to the emergence of two gapped energy bands such that excitations occur vertically in quasimomentum space from the lower to the upper band.
As for the lattice modulation spectroscopy we develop a tight-binding description for both situations. 

\subsection{The response of a homogeneous optical lattice}
\begin{figure}
\includegraphics[width=.99\columnwidth,clip=true]{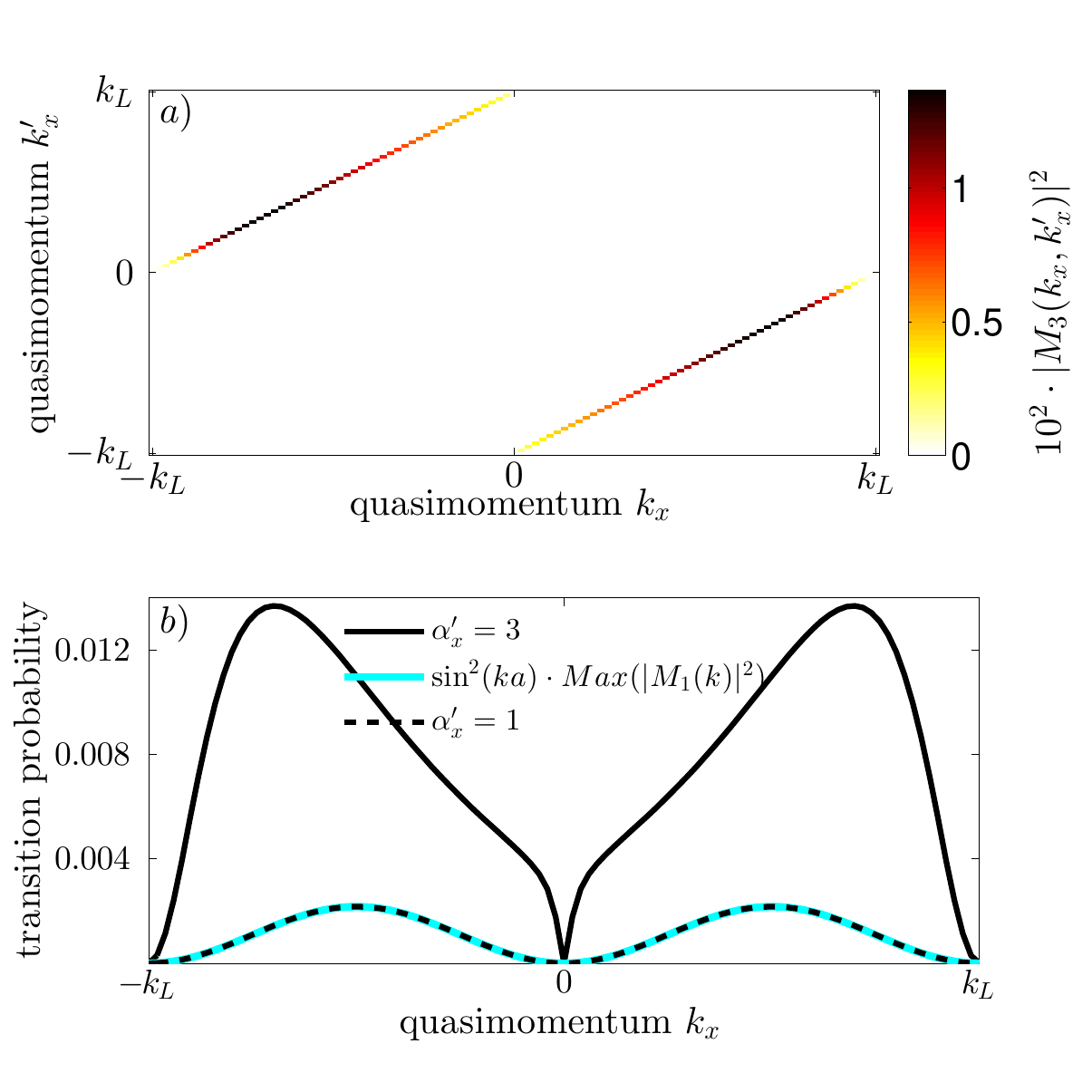}
 \caption{a) The transition probability $\vert M_{(\alpha=1,\vec{k})\rightarrow (\alpha'=3,\vec{k}')}\vert^{2}$ for exciting atoms to  the second excited band $\alpha'=3$ for all $k_x'$ by superlattice modulation. b)  The transition probability $\vert M_{\alpha_x'}(k_x)\vert^2$ for exciting atoms from the lowest band $\alpha=1$ at quasimomentum $k_x$ to the lowest band $\alpha'=1$ or second excited band $\alpha'=3$ at quasimomentum $k_x'=k_x+k_L$.}
\label{fig:fig6}
\end{figure}
\begin{figure}
\includegraphics[width=.99\columnwidth,clip=true]{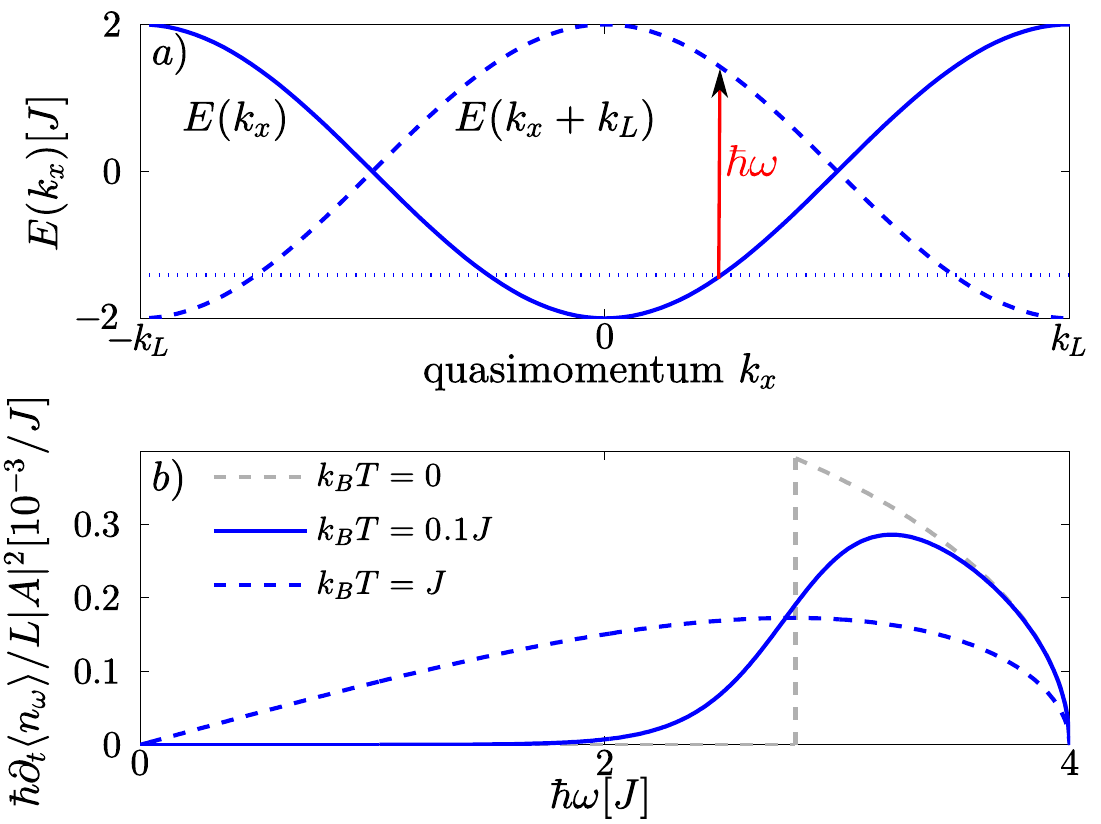}
 \caption{a) A superlattice perturbation can lead to excitations within the lowest band $E(k_x)$ with quasimomentum  transfer $\Delta k_x=k_L$. Possible excitations depend on the filling indicated by the chemical potential $\mu$ (dotted line). b) The corresponding temperature-dependent response of the atoms to a superlattice amplitude modulation  for a quarter-filled band and different temperatures.}
\label{fig:fig7}
\end{figure}
\begin{figure}
\includegraphics[width=.99\columnwidth,clip=true]{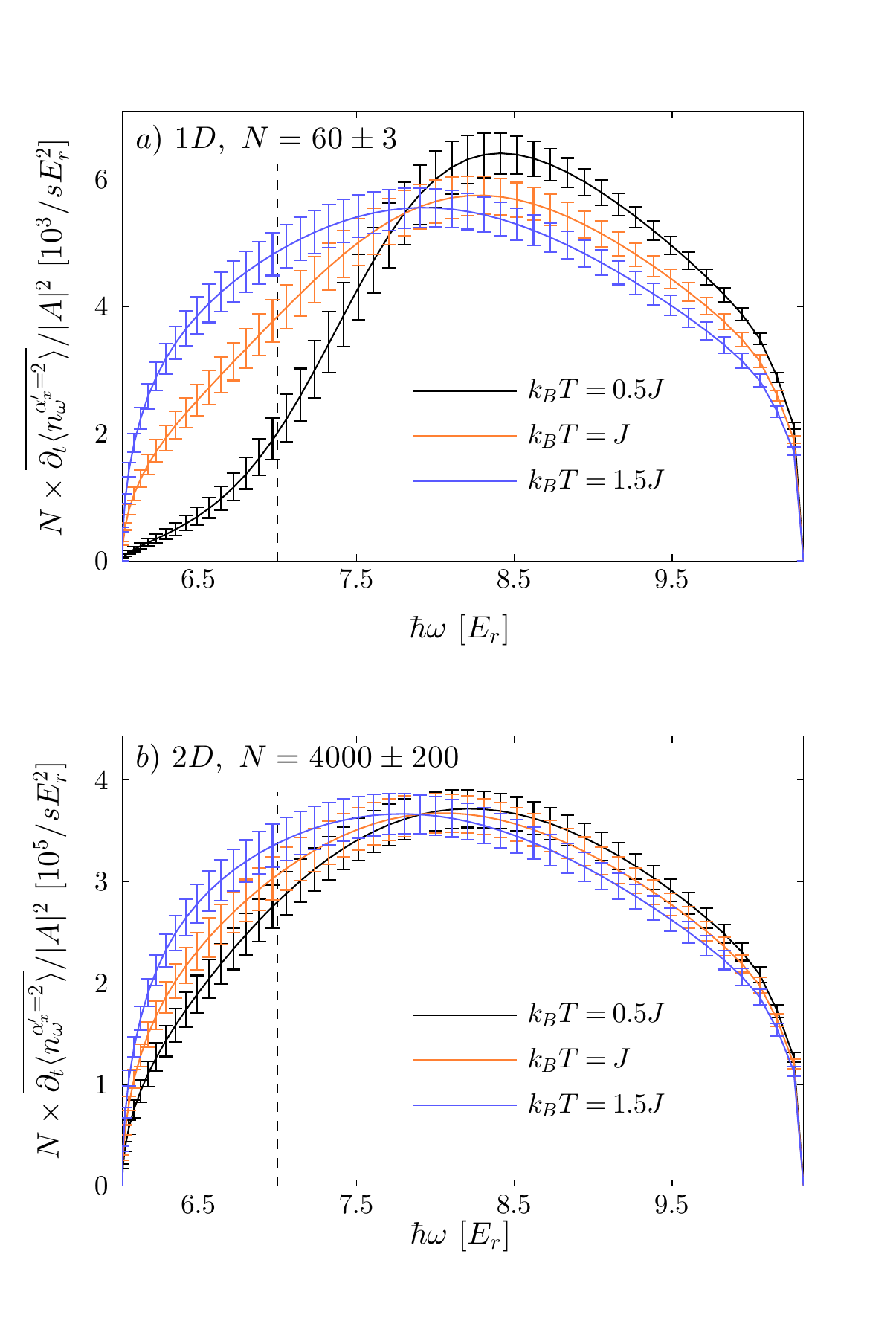}
 \caption{The response to a superlattice modulation along one direction of the trapped system within LDA. In a) the underlying equilibrium lattice is one dimensional with atom number $N=60$. In b) the underlying equilibrium lattice is two dimensional with atom number $N=4000$. In both cases, the response is shown for temperatures $k_BT=0.5$J, $1$J and $1.5$J with error bars corresponding to $5\%$ uncertainty on the initial atom number $N$.}
\label{fig:fig8}
\end{figure}
We start our discussion by the first case, considering an equilibrium system which is given by a homogeneous optical lattice setup and apply a superlattice modulation. We consider the multiple-band model developed in Section \ref{sec:label3}. As before, the equilibrium Hamiltonian $H_0$ is given by Eq. \eqref{eqn:2}. As we consider the superlattice perturbation, we need to compute the transition matrix elements in Bloch basis given by Eq. \eqref{eqn:3} where we now insert the perturbing potential $\delta V(x)\approx \sin(k_Lx)$ which is incommensurate with the underlying equilibrium lattice. We find, that to a good approximation [cf. Fig. \ref{fig:fig6} a) for $\alpha'=3$], the superlattice perturbation only couples quasimomentum $k_x$ in the lowest band to quasimomentum $k_x+k_L$ in higher bands,
\begin{eqnarray}
\nonumber M_{(\alpha=1,\vec{k})\rightarrow (\alpha',\vec{k}')}&=&M_{ \alpha_x'}(k_x,k_x') \prod_{i\neq 1}\delta_{k_{x_i}',k_{x_i}}\\
\nonumber &\approx&M_{ \alpha_x'}(k_x)\delta_{\vec{k}',\vec{k}+(k_L,0,0)}.
\end{eqnarray}
The perturbing part of the Hamiltonian becomes
\begin{eqnarray}
\nonumber H_{\text{pert}}&=&A\sin(\omega t)O,\\
\nonumber O&=&\sum_{\vec{k},\sigma}M_{\alpha_x'}(k_x)\left(c_{\alpha'\vec{k}+(k_L,0,0)\sigma}^{\dag}c_{\alpha=1\vec{k}\sigma}+\text{h.c.}\right).
\end{eqnarray}
In contrast to the commensurate modulation [cf. Eq. \eqref{eqn:4}], transitions within the lowest band are possible due to the finite quasimomentum transfer, provided that empty states at $k_x+k_L$ are available.
The probability $\vert M_{\alpha_x'}(k_x) \vert^{2}$ of exciting atoms within the lowest band $\alpha=1$ to $\alpha'=1$ or from the lowest band $\alpha=1$ to the second excited band $\alpha'=3$  is shown in Fig. \ref{fig:fig6} b). For the probability of excitations within the lowest band we find $\vert M_{ 1}(k_x) \vert^{2}\propto \sin^{2}(k_xa)$. 
 The excitation probability to the second excited band $\vert M_{ 3}(k_x) \vert^{2}$ is similar in symmetry to $\vert M_{ 1}(k_x) \vert^{2}$ but not a pure sinus due to the difference in shape of the excited band. In both cases, prohibited transfers are located at $k_x=0$ and $k_x=k_L$ as for the commensurate modulation when exciting to the first excited band. The response becomes
\begin{eqnarray}
&& \frac{\partial_t\langle n_{\omega}^{\alpha'}\rangle^{dD}}{\vert A\vert^2}=\frac{\pi}{\hbar}\left (\frac{L}{2k_L}\right)^d \int_{BZ}\text{d}^dk\ \vert M_{ \alpha_x'}(k_x)\vert^2 \label{eqn:8}\\
\nonumber  &&\times f\left [E_{\alpha=1}(\vec{k})-\mu\right ] \delta \left\{ \hbar \omega-\left[E_{ \alpha_x'}(k_x+k_L)-E_{\alpha_x=1}(k_x)\right] \right \} ,
\end{eqnarray}
in $d=1$, $2$ or $3$ dimensions.
\\We now discuss the purely one-dimensional model and solely excitations within the lowest band  ($\alpha'=1$).
Atoms with quasimomentum  $k_x$ are excited within the lowest band to an unoccupied state $k_x+k_L$ if available while the system absorbs the energy $\hbar \omega= E(k_x+k_L)-E(k_x)=-2E(k_x)$. We approximate $E(k_x)=-2J\cos(k_xa)$ where $J$ is the tunneling matrix element of the lowest band in $x$-direction. Possible excitations may be represented by transferring atoms at constant quasimomentum $k_x$ to the inverted band $E(k_x+k_L)$ as shown in Fig. \ref{fig:fig7} a). The atom excitation rate [Eq. \eqref{eqn:8}] can be reduced to
\begin{eqnarray}
\nonumber &&\frac{\hbar \partial_t\langle n_{\omega}^{\alpha_x'=1}\rangle}{L\vert A\vert^2}=
-\frac{\text{max}(\vert M_1(k_x)\vert^2)}{4J}\sqrt{1-\left (\frac{\hbar \omega}{4J}\right)}\\
\nonumber &&\times \left[f\left(\frac{\hbar\omega}{2}-\mu\right )-f\left(-\frac{\hbar\omega}{2}-\mu\right)\right].
\end{eqnarray}
The temperature-dependence of the response function is shown in Fig.  \ref{fig:fig7} b). The Fermi tail in the response is stretched by a factor of two with respect to the initial Fermi occupation of the lowest band due to the band inversion. This effect is strongly enhanced by higher-band excitations that hence are in the focus of our discussion.
\\Consequently, in the following we restrict the presentation of results for the atom excitation rate to the second excited band $\alpha'=3$ in the presence of an additional external trapping potential. We study the response to superlattice modulation of the trapped system in the LDA along the lines of Sections \ref{sec:label4} and \ref{sec:label5} choosing the same system parameters and the same initial atom number. The mean atom excitation rate in the trap is given by Eq. \eqref{eqn:5} in 1D and Eq. \eqref{eqn:7} in 2D where we insert the response to the superlattice modulation of the homogeneous system [Eq. \eqref{eqn:8} in 1D and 2D, respectively]. The total atom excitation rate to the second excited band of the 1D and 2D system with $5\%$ uncertainty on the initial atom number is shown in Figs. \ref{fig:fig8} a) and b), respectively.
We observe better frequency resolution compared to the case of commensurate modulation due to the enhanced width of the second excited band. In order to resolve the Fermi tail and distinguish the curves of different temperature in 1D a broad frequency resolution $\Delta (\hbar \omega) \approx \pm 0.15E_r$ is already sufficient in experiment which corresponds to a perturbing time $t\approx 2.5$ms for the chosen parameters. We consider a typical time of $t=10$ms and a perturbation amplitude  $A=0.05V_{0,x}$. At the point  $\hbar \omega=7E_r$ indicated by a dashed line in Fig. \ref{fig:fig8} a) as an example. Using Eq. \eqref{eqn:6} we determine the mean number of excited atoms $\langle n_{\omega}^{\alpha'=3}(t) \rangle$ during time $t$ to lie between $2$ and $6$ for temperatures $k_BT=0.5$J, $1$J and $1.5$J which shows that we excite a couple of hundreds of atoms when considering $\sim 100$ parallel 1D tubes of approximately equal filling with an atom number difference of  $\sim 150$ which is detectable in experiment. In 2D the improved frequency resolution is a clear advantage compared to the commensurate lattice modulation.  We find that a frequency resolution $\Delta (\hbar \omega) \approx \pm 0.025E_r$ is already sufficient in experiment which corresponds to a perturbing time $t\approx 9$ms. Considering an amplitude $A=0.05V_{0,x}$ at $\hbar \omega=7E_r$ we excite atom numbers  $\langle n_{\omega}^{\alpha'=3}(t) \rangle$ between $300$ and $400$ for temperatures $k_BT=0.5$J, $1$J and $1.5$J with an atom number difference of approximately $30$ atoms which is easier to detect in experiment than the smaller atom number difference in the case of commensurate lattice modulation.

\subsection{The response of a dimerized lattice}
\begin{figure}
\includegraphics[width=.99\columnwidth,clip=true]{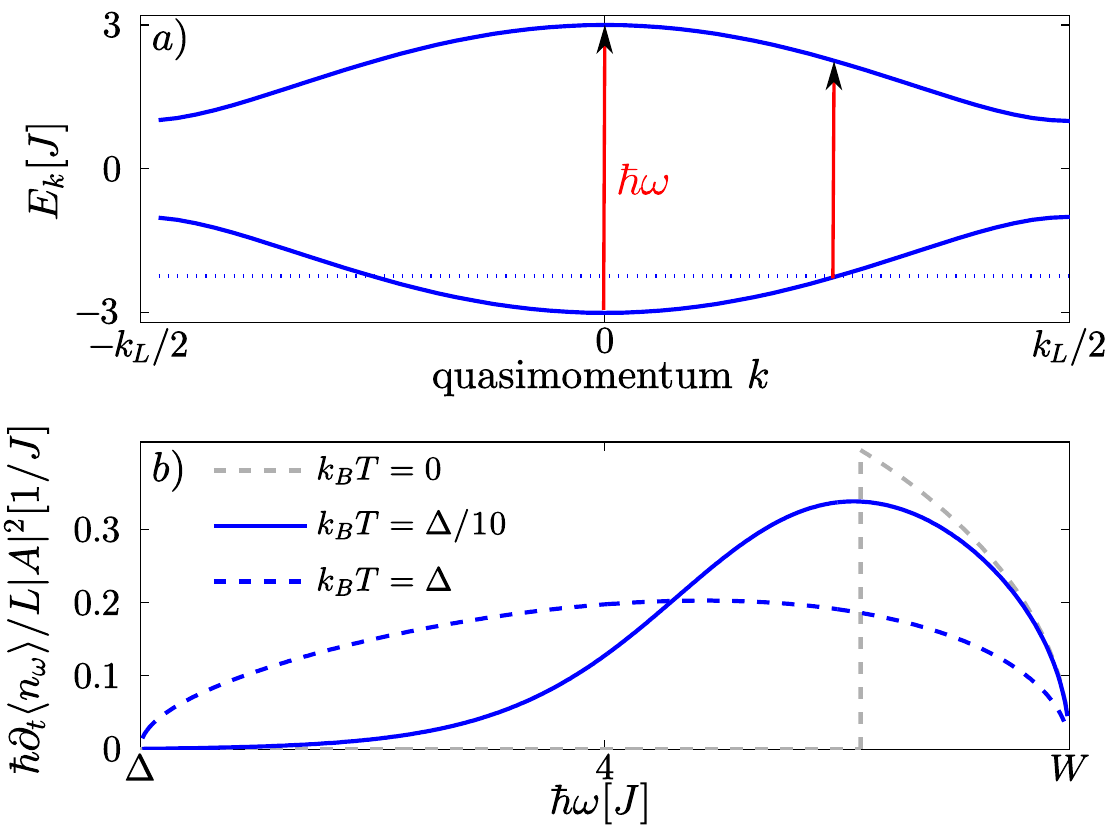}
 \caption{a) The superlattice modulation of the dimerized (superlattice) equilibrium lattice ($J'=2J$) leads to excitations from the lower energy band $-E_k$ to the upper energy band $+E_k$ at conserved quasimomentum $k$. Possible excitations depend on the filling given by the chemical potential $\mu$ (dotted line). b) The corresponding temperature-dependent response function to a superlattice amplitude modulation for a quarter-filled band and different temperatures.}
\label{fig:fig9}
\end{figure}
In this subsection we consider a one-dimensional dimerized system as the equilibrium system and apply the superlattice modulation which has the same geometry as the equilibrium system. The equilibrium superlattice is engineered to have a constant bottom offset but alternating lattice height [cf. Fig. \ref{fig:fig1} c)].
The equilibrium tight-binding Hamiltonian describing the considered superlattice configuration is given by
\begin{eqnarray}
\nonumber H_0&\approx&-J\sum_{j\,\text{odd},\sigma}(c_{j,\sigma}^{\dag}c_{j+1,\sigma}+\text{h.c.})\\
&-&J'\sum_{j\,\text{even},\sigma}(c_{j,\sigma}^{\dag}c_{j+1,\sigma}+ \text{h.c.})-\mu \sum_{j,\sigma}n_{j,\sigma} ,\label{eqn:9}
\end{eqnarray}
where hopping amplitudes $J'>J>0$ alternate between neighboring sites. The unit cell contains two sites that may be assigned to two sub-lattices. Consequently, the first Brillouin zone has half the size $]-k_L/2,k_L/2]$ compared to the homogeneous lattice. We employ a Fourier transform combined with a Bogoliubov transformation  (cf. Appendix \ref{appendix:1} for details) and obtain
\begin{eqnarray}
\nonumber H_0&=&\sum_{k,\sigma}E_k(\alpha_{k,\sigma}^{\dag}\alpha_{k,\sigma}-\beta_{k,\sigma}^{\dag}\beta_{k,\sigma})\\
&-&\mu\sum_{k,\sigma}(\alpha_{k,\sigma}^{\dag}\alpha_{k,\sigma}+\beta_{k,\sigma}^{\dag}\beta_{k,\sigma}), \label{eqn:10}
\end{eqnarray}
which leads to the emergence of two energy bands $\pm E_k=\pm \sqrt{J'^2+J^2+2JJ'\cos(2ka)}$ gapped by an energy $\Delta=2(J'-J)$ and of total band width $W= 2(J'+J)$ [cf. Fig. \ref{fig:fig9} a)]. Here $k=k_x$ denotes the quasimomentum in $x$-direction. The operators $\alpha_{k,\sigma}^{(\dag)}$  and $ \beta_{k,\sigma}^{(\dag)}$ annihilate (create) quasiparticles in the upper or lower band, respectively.
Excitations due to the superlattice modulation occur from the lower band $- E_k$ to the upper band $+ E_k$ at conserved quasimomentum $k$ as shown in Fig. \ref{fig:fig9} a). This is in contrast to the case of a homogeneous equilibrium lattice where the superlattice modulation introduces quasimomentum $\Delta k=k_L$ into the system. However, here, quasimomentum is conserved as the equilibrium lattice is already dimerized and $\Delta k=k_L=0$ in the reduced zone scheme of the superlattice. For the atom excitation rate to the upper band we obtain (see Appendix \ref{appendix:1} for details)
\begin{eqnarray}
 \frac{\hbar \partial_t\langle n_{\omega}^{\text{upper band}}\rangle}{L\vert A\vert^2}&=&\frac{\vert C(\omega) \vert ^2F(\omega,\beta,\mu)}{\vert g'(\omega)\vert },\label{eqn:11}
\end{eqnarray}
where the functions $ \vert C(\omega)\vert^2$ and $\vert g'(\omega)\vert$ do not depend on temperature (cf. Appendix \ref{appendix:1}) and the temperature-dependent part is given by 
\begin{eqnarray}
\nonumber F(\omega,\beta,\mu)&=&\left ( \frac{e^{\beta\omega/2}-e^{-\beta\omega/2}}{e^{-\beta \mu}(e^{\beta \mu}+e^{\beta\omega/2})(e^{\beta \mu}+e^{-\beta\omega/2})}\right ).
\end{eqnarray}
The strong temperature-dependence of the response is shown in Fig.  \ref{fig:fig9} b). The temperature-dependent tail is also stretched by a factor two in energy as for the homogeneous equilibrium lattice when exciting within the lowest band but excited states are now clearly separated from the initial distribution which is an advantage. Consequently, a simple thermometry scheme is available for the dimerized equilibrium lattice by superlattice modulation spectroscopy by counting the number of excited atoms in the upper band  if the parameters allow for sufficient frequency resolution. Here, we do not consider explicit experimental parameters as our aim was to outline the conceptual scheme in in the dimerized superlattice. However, it can be extended, incorporating typical experimental parameters.

\section{Conclusion}
In this work we have studied a non-interacting Fermi gas confined to optical lattices. We probed the response to a time-periodic modulation of the lattice amplitude. We investigated the temperature-dependent atom excitation rate to higher Bloch bands and demonstrated that it shows clear signatures of the Fermi distribution of the equilibrium system. We explored the possibilities of thermometry for different geometries of the equilibrium system as well as for different amplitude modulation schemes, either commensurate or incommensurate with the equilibrium system.\\We first considered a homogeneous equilibrium lattice subjected to both a homogeneous lattice amplitude modulation [Fig. \ref{fig:fig1} a)] and a lattice amplitude modulation of superlattice geometry [Fig. \ref{fig:fig1} b)]. We studied the excitations to higher Bloch bands on the basis of a multiple-band model which we constructed for typical experimental parameters. We found that for the homogeneous modulation quasimomentum is conserved and excitations to the first excited band are suitable for thermometry. In contrast, the superlattice modulation transfers quasimomentum $k_L$ into the system and we consider excitations to the second excited band. In both cases, the response displays a clear signature of the temperature-dependent Fermi distribution in the lowest band for one -and two-dimensional equilibrium systems where we considered temperatures on the order to a few percent of the hopping amplitude $J$. The Fermi distribution is strongly broadened in energy due to the much larger bandwidth of the higher bands compared to the lowest band. We benefit from this because the experimental frequency resolution required to resolve the Fermi dependence is achievable within typical durations of the perturbation.  We established close relation to experiments by incorporating an external trapping potential and by estimating the number of excited atoms in order to verify the measurability. Additionally, we investigated the response of a dimerized equilibrium system subjected to the superlattice modulation setup [Fig. \ref{fig:fig1} c)]. We found that this setup is also suitable for thermometry due to the emergent two-band structure.\\We emphasize that the temperature-dependence of the atom excitation rate becomes more distinct for decreasing temperatures such that our scheme extends down to temperatures that have not been reached in experiment so far and covers the regime of interest below the N\'eel temperature where antiferromagnetic ordering is expected to occur.\\Prospectively, it is of interest to incorporate interactions which will require more sophisticated theory but shall not change the basic mechanism of the measurement scheme. \\The superlattice modulation applied to a homogeneous system injects non-zero quasimomentum  $k_L$ into the system. Quite generally, one may realize arbitrary quasimomentum  transfer $K$ if choosing different geometry of the perturbation, i.e. replacing the dimerization $(-1)^j$ in Eq. \eqref{eqn:app1} by $\cos(Kj)$. This is promising as a spectroscopic probe as it allows for the investigation of more complex lattice models and various intriguing quantum phases.

\begin{acknowledgments}
We thank S. Wolff, A. Rosch and the group of M. K\"ohl, in particular L. Miller, J. Drewes and F. Brennecke, for insightful discussions. We acknowledge DFG (BCGS) for financial support.
\end{acknowledgments}

\appendix
\section{The superlattice modulation of the dimerized equilibrium lattice}\label{appendix:1}
We diagonalize the dimerized tight-binding Hamiltonian given by Eq. \eqref{eqn:9}. First, we assign two sub-lattices with annihilation (creation) operators on odd sites and even sites defined by
\begin{eqnarray}
 \nonumber a_{j\sigma}^{(\dag)}&=&c_{2j+1,\sigma}^{(\dag)},\\
\nonumber b_{j\sigma}^{(\dag)}&=&c_{2j,\sigma}^{(\dag)}.
\end{eqnarray}
Note that the first Brillouin zone is divided in half $]-k_L/2,k_L/2]$ compared to the homogeneous equilibrium lattice.
We then employ the Fourier transform,
\begin{eqnarray}
\nonumber a_{j,\sigma}&=& \sqrt{\frac{2}{L}}\sum_ke^{ik2aj}a_{k,\sigma},\\
\nonumber b_{j,\sigma}&=& \sqrt{\frac{2}{L}}\sum_ke^{ik2aj}b_{k,\sigma},
\end{eqnarray}
where $k=k_x$, combined with a Bogoliubov transformation,
\begin{eqnarray}
\nonumber  \alpha_{k,\sigma}&=&u_ka_{k,\sigma}-v_kb_{k,\sigma},\\
\nonumber \beta_{k,\sigma}&=&u_ka_{k,\sigma}+v_kb_{k,\sigma},
\end{eqnarray}
 where
\begin{eqnarray}
\nonumber u_k&=&\frac{1}{\sqrt{2}}\exp \left [-i\frac{\varphi(k)-ka}{2}\right ],\\
\nonumber v_k&=&\frac{1}{\sqrt{2}}\exp \left [i\frac{\varphi(k)-ka}{2}\right ],
\end{eqnarray}
with
\begin{eqnarray}
\nonumber \varphi (k)= \begin{cases} \arctan \left [-\frac{(J'-J)\sin(ka)}{(J'+J)\cos(ka)}\right ],\qquad k\in \left(-\frac{k_L}{2},\frac{k_L}{2}\right)\\
  -\frac{\pi}{2},\qquad k=\frac{k_L}{2}\end{cases}.
\end{eqnarray}
With this we obtain the diagonal equilibrium Hamiltonian Eq. \eqref{eqn:10} of the superlattice which exhibits two energy bands $\pm E_k=\pm \sqrt{J'^2+J^2+2JJ'\cos(2ka)}$. We approximate the superlattice perturbation by the dimerization operator
\begin{eqnarray}
O&=&\sum_{j,\sigma} (-1)^j\left ( c_{j,\sigma}^{\dag} c_{j+1,\sigma} +\text{h.c.} \right )\label{eqn:app1}.
\end{eqnarray}
Using the above transformation we obtain
\begin{eqnarray}
\nonumber O&=&\sum_{k,\sigma}C_0(k)\left(\alpha_{k,\sigma}^{\dag}\alpha_{k,\sigma}-\beta_{k,\sigma}^{\dag}\beta_{k,\sigma}\right)+\\
\nonumber &+&C(k)\left(\beta_{k,\sigma}^{\dag}\alpha_{k,\sigma}-\text{h.c.}\right),
\end{eqnarray}
with
\begin{eqnarray}
\nonumber C_0(k)&=&2\sin(ka)\sin[\varphi (k)],\\
\nonumber C(k)&=&2i\sin(ka)\cos[\varphi (k)].
\end{eqnarray}
The resulting atom excitation rate to the upper band is given by Eq. \eqref{eqn:11} where
\begin{eqnarray}
\nonumber \vert C(\omega)\vert^2&=&\omega^2\left [-\frac{(J+J')^2}{16J^2J'^2}\right ]+\frac{(J+J')^2(J^2+J'^2)}{2J^2J'^2}\\
\nonumber &+&\frac{1}{\omega^2}\left [ -\frac{(J+J')^2(J^2-J'^2)^2}{J^2J'^2}\right ],\\
\nonumber \vert g'(\omega)\vert&=&\frac{4JJ'\sqrt{1-\frac{[\omega^2-4(J^2+J'^2)]^2}{64J^2J'^2}}}{\omega/2}.
\end{eqnarray}
\\
\section{A comparison to exact diagonalization}\label{app:comparison}

In order to verify that the LDA approximation is justified, we have treated the trapping potential exactly. We here concentrate on the case of a 1D system subjected to the commensurate lattice modulation.

The simulations are performed using exact diagonalization on
the discretized version of the single particle Hamiltonian in the presence of the lattice and the harmonic trap. The space is discretized using $\Delta_x=a/50$ in $x$-direction considering 500 wells in the lattice.

The atom excitation rate from the lowest band to the excited bands is calculated using Eqs.~\eqref{eqn:3} and \eqref{eqn:5a} in the discrete form 
\begin{eqnarray}
&&M_{(\alpha_x=1,i)\rightarrow (\alpha_x',i')}=\langle v_{\alpha_x',i'}|\delta V_x|v_{\alpha_x,i}\rangle,\nonumber\\
&&N\frac{1}{\vert A\vert^2}\partial_t\langle n_{\omega}^{\alpha_x'}\rangle^{1D}=\frac{\pi}{\hbar} \sum_{i,i'} \vert M_{(\alpha_x=1,i)\rightarrow (\alpha_x',i')} \vert^2 \nonumber\\
&&\times f\left [E_{\alpha_x=1}(i)-\mu)\right ] \frac{1}{\sqrt{2\pi w^2}} e^{-\frac{1}{2\sigma^2}\left(\hbar \omega-\left[E_{ \alpha_x'}(i')-E_{\alpha_x=1}(i)\right]\right)^2}\nonumber,
\label{eqn:B3}
\end{eqnarray}
where $E_{\alpha_x}(i)$ and $|v_{\alpha_x,i}\rangle$ are the eigenenergy and the eigenstate for energy band $\alpha_x$ calculated from the exact diagonalization and $i$ labels different states in each band.
$\delta V_x$ is the perturbing potential in the discrete form and the Dirac delta function is replaced by its Gaussian approximation with width $w\approx 0.007 E_{r}$. 
 \begin{figure}
\includegraphics[width=.99\columnwidth,clip=true]{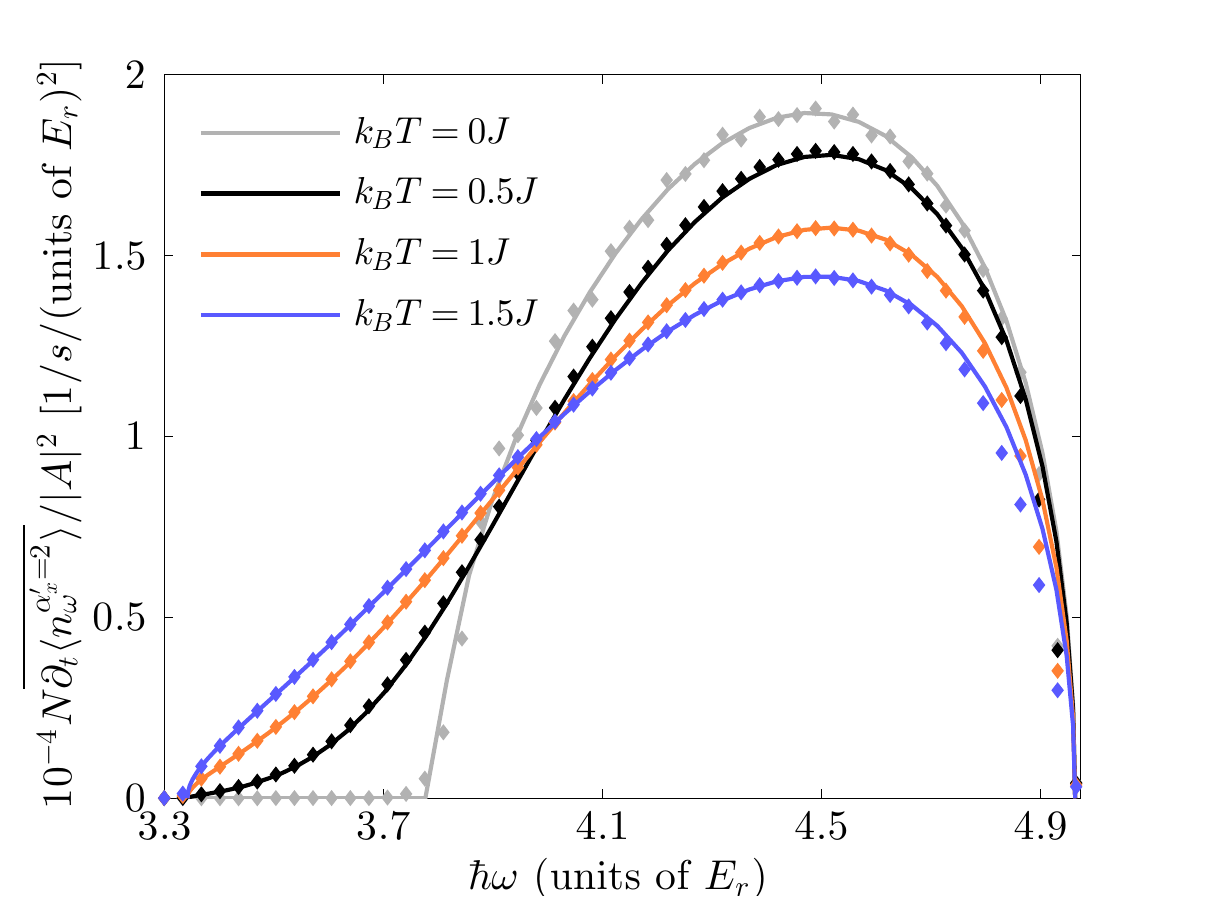}
 \caption{Atom excitation rate to the first excited band $\alpha_x=2$ induced by the commensurate lattice modulation spectroscopy of the trapped system calculated in LDA (solid lines) and by exact diagonalization (diamonds). We consider the one-dimensional equilibrium system with initial atom number $N=60$. The response is shown for temperatures $k_BT=0$, $0.5$J, $1$J and $1.5$J.}
\label{fig:figA1}
\end{figure}
A comparison of the LDA and exact diagonalization results for the atom excitation rate to the first excited band for all temperatures considered in the main part of this work is shown in Fig. \ref{fig:figA1}. We find excellent agreement which justifies the use of LDA.

\bibliographystyle{h-physrev}

\end{document}